\documentclass[amssymb]{article}
\usepackage{graphicx}
\usepackage{amsfonts}

\newcommand{\be}{\begin{equation}}
\newcommand{\ee}{\end{equation}}
\newcommand{\bea}{\begin{array}}
\newcommand{\ea}{\end{array}}
\newcommand{\beqa}{\begin{eqnarray}}
\newcommand{\eeqa}{\end{eqnarray}}
\newcommand{\bean}{\begin{eqnarray*}}
\newcommand{\eean}{\end{eqnarray*}}

\def\up#1{\leavevmode \raise.16ex\hbox{#1}}

\newcommand{\gapproxeq}{\lower
 .7ex\hbox{$\;\stackrel{\textstyle >}{\sim}\;$}}
\newcommand{\lapproxeq}{\lower .7ex\hbox{$\;\stackrel
{\textstyle <}{\sim}\;$}}

\newcounter{appendice}

\def\thebibliography#1{{\bf REFERENCES\markboth
 {REFERENCES}{REFERENCES}}\list
 {[\arabic{enumi}]}{\settowidth\labelwidth{[#1]}\leftmargin\labelwidth
 \advance\leftmargin\labelsep
 \usecounter{enumi}}
 \def\newblock{\hskip .11em plus .33em minus -.07em}
 \sloppy
 \sfcode`\.=1000\relax}

\def\BI{{\rm 1\!l}}

\begin{document}

\centerline{ \LARGE Fuzzy $CP^2$ Space-Times }

\vskip 2cm
\centerline{A. Chaney\footnote{adchaney@crimson.ua.edu} and A. Stern\footnote{astern@ua.edu} }
\vskip 1cm
\begin{center}
{ Department of Physics, University of Alabama,\\ Tuscaloosa,
Alabama 35487, USA\\}
\end{center}
\vskip 2cm
\vspace*{5mm}
\normalsize
\centerline{\bf ABSTRACT}

Four-dimensional manifolds with changing signature are obtained by taking  the large $N$ limit of fuzzy $CP^2$ solutions to a Lorentzian matrix model.  The regions of  Lorentzian signature give toy models of closed  universes which   exhibit  cosmological singularities.   These singularities are resolved at finite $N$, as the underlying  $CP^2$ solutions are expressed in terms of  finite matrix elements.
\bigskip
\bigskip

\newpage

\section{Introduction}
Fuzzy spheres are defined by  $N\times N$ irreducible matrix representations of the $su(2)$ algebra.\cite{Madore:1991bw}-\cite{Iso:2001mg}
In a previous work,\cite{Chaney:2015mfa} we showed that a fuzzy sphere embedded in a   Minkowski background, which we denote by $S^{2,L}_{F}$, can serve as a two-dimensional toy model of a closed noncommutative cosmology.  Noncommutative or matrix cosmologies have been of interest for some time, and they  possess  a limit, the `commutative' limit, where a space-time manifold is recovered from the matrix configurations.\cite{Alvarez:1997fy}-\cite{Stern:2014aqa} 
  In \cite{Chaney:2015mfa}, the commutative limit for  $S^{2,L}_{F}$ corresponds to taking  $N\rightarrow \infty$, which yields  a sphere embedded in  Minkowski space.  This `sphere'  had several novel features.   The  curvature  computed from the induced metric is not  constant and there are singularities at two fixed latitudes.  Also, the induced metric has changing  signature. Signature change is known to be a possible feature of both classical and quantum gravity.\cite{Sakharov:1984ir}-\cite{Ambjorn:2015qja}  The  region bounded by the singular latitudes has  Lorentzian signature, and describes a closed two-dimensional space-time.     The two singular latitudes behave as cosmological singularities, which get resolved at finite $N$. 

The question naturally  arises as to whether one  can generalize  $S^{2,L}_{F}$ to four-dimensional fuzzy cosmologies. Of course, a trivial generalization is obtained by taking the tensor product of two noncommutative spaces, for example  $S^{2,L}_{F}\times S^{2}_{F}$ ,  $S^{2}_{F}$ being a fuzzy sphere in a Euclidean background.  Such tensor product spaces appear after extremizing  the sum of two bosonic
 matrix actions, consisting of  Yang-Mills terms, analogous to what appears in the Ishibashi, Kawai, Kitazawa, Tsuchiya (IKKT) model,\cite{Ishibashi:1996xs}   along with cubic terms and mass terms.  The results of \cite{Chaney:2015mfa} can be straightforwardly repeated in this case.
 
   Here, instead, we examine fuzzy $CP^2$ ($CP^2_F$).\cite{Balachandran:2005ew},\cite{Alexanian:2001qj}-\cite{Ydri:2016dmy} In order for time to emerge in the large $N$ limit we embed $CP^2_F$  in a Lorentzian background.  This $CP^2_F$  results from extremizing a matrix model action, again consisting of a  Yang-Mills term, cubic term and mass term.  The large $N$ limit yields four dimensional
manifolds which are, loosely speaking, embeddings of $CP^2$  in an eight (or greater)-dimensional Lorentzian target space with a flat metric tensor.  Analogous to the  two-dimensional model in \cite{Chaney:2015mfa},
the induced  metric tensor on the four-dimensional surface can have changing signature.  Signature changes occur at two three-dimensional singular surfaces, which define the boundaries between regions of Euclidean and Lorentzian signature. 
 The region of Lorentzian signature defines a closed  space-time, with the singular surfaces playing the role of  cosmological singularities.  A novel feature of these toy universes is that the cosmological singularities occur at nonzero distance scales, and that time cannot be defined for smaller distance scales. As with the two-dimensional models in \cite{Chaney:2015mfa}, the singularities appear only after taking the large $N$ limit, and so the finite $N$ matrix description once again resolves the singularities of the continuum description.

The outline of this article is the following:
We review $CP^2$ in section two and $CP^2_F$ in section three.  In section four we show that  $CP^2_F$  solutions result from both Euclidean and Lorentzian Yang-Mills type-matrix models. 
A one-parameter family of deformed  $CP^2_F$  solutions (which contains the undeformed solution)  is also found to Lorentzian Yang-Mills type-matrix model,  which requires a mass term, along with a cubic term, in the action.  The large $N$ (or commutative) limit of these solutions is taken in section five.  There we plot the distance scale versus time in the comoving frame for the  $CP^2$ universes. Concluding remarks are given in section six.

\section{$CP^2$} 
To define
$CP^2$ one starts with a three-dimensional complex vector space spanned by  $z=(z_1,z_2,z_3)$, where $z_i\in C$ are not all zero,  and then makes the identification of $z$ with  $ \gamma z$, for all complex nonzero values of $\gamma$.  
$CP^2$ can equivalently be defined as the space of $U(1)$ orbits on  the unit $5-$sphere $S^5$.  The latter is spanned by $z$ with  $|z|^2=z^*_iz_i=1$, where $i$ is summed from $1$ to $3$, and a point on the space of $U(1)$ orbits is  $\{e^{i\beta} z,\;0\le\beta<2\pi\}$.  Upon introducing the following Poisson brackets
\be \{z_i, z^*_
j \} =-i\delta_{ij}\qquad\;\; \{z_i, z_
j \} =  \{z^*_i, z^*_
j \} = 0\;,\label{pbsfrzzstr}\ee one can generate the $U(1)$ orbits from  the  $5-$sphere constraint
\be {\cal C}= z^*_iz_i-1\approx 0\;\label{sfive}\ee
Infinitesimal variations $\delta_\epsilon$ along an orbit are then
\be  \delta_\epsilon z_i= \{z_i, {\cal C}\}\epsilon= -i\epsilon z_i  \qquad\quad \delta_\epsilon z_i^*= \{z_i^*, {\cal C}\}\epsilon= i\epsilon z_i ^* \;,\ee where $\epsilon$ is an infinitesimal parameter.

$CP^2$ is also defined as $SU(3)/U(2)$, i.e., the space of adjoint  orbits of $SU(3)$ through
$\lambda^8$, where $\lambda^\alpha$, $\alpha=1,2,...,8$ are  the Gell-Mann matrices, i.e., $CP^2=\{U\lambda^8 U^\dagger, U\in SU(3)\}$.  Upon introducing 
\be x^\alpha=\frac {\bar z \lambda^\alpha z} {|z|^2}\;,\label{cmsu3gen}\ee 
one recovers the    $su(3)$ Lie algebra 
from the Poisson bracket algebra 
 on $CP^2$. 
 Using the commutator   $[\lambda^\alpha,\lambda^\beta] = 2if^{\alpha\beta\gamma}\lambda^{\gamma}$ we get from (\ref{pbsfrzzstr})   that
 \be \{x^\alpha,x^\beta\} =\frac {2}{|z|^2}f^{\alpha\beta\gamma}x^{\gamma}\;,\label{pbsfrmbdcrds}\ee
after imposing the constraint (\ref{sfive}).  $x^\alpha$ are functions of $z$ and $\bar z$ which are invariant under $z\leftrightarrow \gamma z$, $\gamma\in C$,
and so they span a four-dimensional constrained surface, i.e., $CP^2$, in $ {\mathbb{R}}^8$. The constraints on $x^\alpha$ are 
 \be  x^\alpha x^\alpha = \frac 43\qquad\qquad d^{\alpha\beta\gamma} x^\beta x^\gamma  =\frac 13 x^\alpha \;, \label{cntuumvls} \ee where $d^{\alpha\beta\gamma}$ are defined from the anticommutator of Gell-Mann matrices  $[\lambda^\alpha,\lambda^\beta] _+=  \frac 43  \delta_{\alpha,\beta}\BI_3 +2d^{\alpha\beta\gamma}\lambda^\gamma$, $\BI_3$ being the $3\times 3$ identity matrix. The constraints in 
 (\ref{cntuumvls}) follow from the expression for $x^\alpha$ in (\ref{cmsu3gen}).

The metric on $CP^2$ is  given by
\be ds_E^2=\frac 4{|z|^4}\Bigl(|z|^2 |dz|^2-|\bar z dz|^2\Bigr) \;, \label{FSm}\;\ee where    $|d z|^2= dz^*_idz_i$ and $\bar z dz=z^*_idz_i$.  It is known as the  Fubini-Study metric and is  invariant under: $z\rightarrow \gamma z$,    $dz\rightarrow d\gamma\, z+ \gamma dz $.  The isometry of the metric is $SU(3)/Z_3$, with corresponding transformations: $z\rightarrow Uz,\; U\in SU(3)/Z_3$.
The Fubini-Study metric is recovered from the embedding    (\ref{cmsu3gen}) of $CP^2$ in the ${\mathbb{R}^8}$ target space, where one assumes a flat Euclidean metric tensor.  That is, starting with the $SO(8)$ invariant
\be  ds_E^2= dx^\alpha dx^\alpha\;, \label{atedemtrc}\ee  and then substituting (\ref{cmsu3gen}),
one recovers  (\ref{FSm}). 

The metric tensor in (\ref{FSm}) can be re-expressed in terms of   a pair of complex coordinates
 $\zeta_a= z_a/z_3$, $a=1,2$ (away from $z_3=0$), which are invariant under $z\rightarrow \gamma z$. 
  Along with their complex conjugates, they span $CP^2$ when $z_3\ne 0$.
 In terms of these coordinates, the invariant length (\ref{FSm}) becomes
\be ds_E^2=2\,g_{a\bar b}\, d\zeta_a d \zeta_b^*=\frac 4{(|\zeta|^2+1)^2}\Bigl((|\zeta|^2+1) |d\zeta|^2-|\bar \zeta d\zeta|^2\Bigr)  \label{FSm2}\;,\ee where $|\zeta|^2=\zeta_a^*\zeta_a$,  $|d\zeta|^2=d\zeta_a^*d\zeta_a$ and $\bar \zeta d\zeta=\zeta_a^*d\zeta_a$.  It is well known to satisfy the sourceless Einstein equations with a positive cosmological constant, specifically $\Lambda=\frac 32$.  From (\ref{pbsfrzzstr}), the Poisson brackets are given by
\be \{\zeta_a, \zeta^*_
b \} =-i(|\zeta|^2+1)(\zeta_a\zeta_b^*+\delta_{ab})\qquad\;\; \{\zeta_a, \zeta_
b \} =  \{\zeta^*_a, \zeta^*_
b \} = 0\;,\label{pbsfrzeta}\ee 
 Their inverse gives the symplectic two-form 
\be \Omega=-\frac i 2 \,g_{a\bar b}\, d\zeta_a \wedge d \zeta_b^*\label{klr2frm} \;,\ee which is also the K\"ahler two-form.

The invariant length and   K\"ahler two-form can furthermore be  expressed in terms of left-invariant Maurer-Cartan one forms $\omega_i$ on $SU(2)$, satisfying $d\omega_i+\epsilon_{ijk}\omega_j\wedge\omega_k=0$.  For this take $\omega_i=\frac i2$Tr$\,\sigma_i\, u^\dagger du$, where $u$ is the $SU(2)$ matrix
\be u=\frac 1{|\zeta|}\;\pmatrix{\zeta_1^*&-i\zeta_2 \cr -i\zeta_2^*  &\zeta_1 }\label{Oilerngls}\ee  and $\sigma_i$ are Pauli matrices. One can write
\be ds_E^2=\frac{4 (d|\zeta|)^2}{(|\zeta|^2+1)^2}  +\frac{4 |\zeta|^2}{(|\zeta|^2+1)}\,(\omega_1^2+\omega_2^2)+\frac{4 |\zeta|^2}{(|\zeta|^2+1)^2}\,\omega_3^2\;\label{cp2mtrcomga}\ee 
and
\be \Omega=-2\, d\biggl(\frac{ |\zeta|^2}{(|\zeta|^2+1)}\,\omega_3\biggr) \label{smplct2frm3} \ee

The isometry of the metric tensor is $SU(3)/Z_3$.  For any $|\zeta|\,(\ne 0)\,$-slice, the metric tensor and  symplectic two-form  are invariant under $SU(2)\times U(1)/Z_2$.  The latter symmetry is also present for the manifolds we obtain in section five. (On the other hand, the $SU(3)/Z_3$ isometry is broken for those manifolds.) The  $SU(2)\times U(1)/Z_2$ transformations on $u$ are of the form $u\rightarrow u'=v\,u\,e^{i\lambda\sigma_3}$, $v\in SU(2)$ and $\lambda\in R$, which leave $\omega_3$ and $\omega_1^2+\omega_2^2$ invariant.  We can parametrize the $SU(2)$ matrices in  (\ref{Oilerngls}) by Euler angles $(\theta,\phi,\psi)$ according to 
\be  \zeta_1  = e^{i\bigl(\frac {\psi+\phi}2\bigr)} \cos{\frac \theta 2}\;  |\zeta|   \qquad\quad \zeta_2= e^{i\bigl(\frac {\psi-\phi}2\bigr)} \sin{\frac \theta 2}\;|\zeta|\;, \label{EFchrt}\ee where in order to span all of $SU(2)$, $0\le \theta <\pi\;$,  $\;0\le \psi<2\pi$  and $\;0\le \phi\le 4\pi$. On the other hand, to parametrize  the  Maurer-Cartan one forms, we only need $\phi$ to run from $0$ to $2\pi$. In terms of the Euler angles, the metric is given by
\be ds_E^2=\frac{4 (d|\zeta|)^2}{(|\zeta|^2+1)^2}  +\frac{|\zeta|^2}{(|\zeta|^2+1)} \,(d\theta^2+\sin^2\theta d\phi^2)+\frac{|\zeta| ^2}{(|\zeta|^2+1)^2}\,(d\psi+\cos\theta d\phi)^2\;\ee
 The Killing vectors $k_a,\; a=1,...,4$, generating the $SU(2)\times U(1)/Z_2$  isometry  group  on  any $|\zeta|\,(\ne 0)\,$-slice  are expressed in terms of the Euler angles according to 
\be
 k_1\pm i k_2= e^{\pm i\phi} \biggl(\frac\partial{\partial\theta}\pm i\Bigl(\cot\theta\frac\partial{\partial\phi}-\csc\theta\frac\partial{\partial\psi}\Bigr)\biggr)
\qquad\quad
k_3=\frac\partial{\partial\phi}\qquad\quad  k_4=\frac\partial{\partial\psi}\;\label{klngvctrs}
\ee

In subsection 5.1 we shall replace the eight-dimensional Euclidean target space by an eight-dimensional Minkowski space to get an alternative metric on `$CP^2$'.  Before doing this we first review $CP^2_F$, the fuzzy analogue of $CP^2$ in section three.

\section{ $CP^2_F$}
\setcounter{equation}{0}

Loosely speaking, $CP_F^2$ is  the quantization of $CP^2$.  For this one replaces the complex coordinates
$z_i$ and $z_i^*$, $ i = 1, 2, 3,$  by  operators   $a_i^\dagger$ and $a_
i$,\cite{Balachandran:2005ew},\cite{Schneiderbauer:2016wub} satisfying the commutation relations of raising and lowering operators
 \be [a_i, a^\dagger_
j ] =\delta_{ij}\qquad\;\; [a_i, a_j ] = [a^\dagger_
i , a^\dagger_j ] = 0\;,\label{hocrs}\ee
acting on some Hilbert space  ${\cal  H}_n$.
 In analogy with  the $5-$sphere constraint
(\ref{sfive}), one fixes the eigenvalue of the total number operator  $a^\dagger_i a_i$ to some positive integer value $n$.
This  restricts 
${\cal  H}_n$ to be spanned by  the  $N=\frac{ (n+2)!}{
2 n!}\;$ harmonic oscillator states $ \{|n_1, n_2, n_3 >\}$,  $n_i=0,1,2,...$, where $n = n_1 + n_2 + n_3$.  The action of   the raising and lowering  operators  is incompatible with this restriction, so $a_i^\dagger$ and $a_
i$  cannot generate the algebra of $CP_F^2$.  One can instead work with functions of the operators $a^\dagger_i a_j$, which do have a well defined action on the  $N-$dimensional Hilbert space  ${\cal  H}_n$. Of course, $a^\dagger_i a_i$ acts trivially on  ${\cal  H}_n$.  The remaining operators, $a^\dagger_i a_j-\frac 13\delta_{ij} a^\dagger_k a_k $, generate $SU(3)$ and are  the noncommutative analogues of (\ref{cmsu3gen}), which we can also write as
\be X^\alpha =\frac 1na^\dagger_
i\lambda^\alpha _{ij}a_j \;,\qquad\;\; \alpha= 1,2,..., 8 \;\label{dffzeecp2}\ee
From them we  recover the $su(3)$ Lie-algebra
\be [X^\alpha,X^\beta] =\frac {2i}nf^{\alpha\beta\gamma}X^{\gamma}\label{criaib}\ee
 $X^\alpha$ acting on ${\cal  H}_n$ generate an irreducible  representation of $ SU(3)$, which is uniquely specified by the values of the quadratic and cubic Casimirs, $X^\alpha X^\alpha$ and $d^{\alpha\beta\gamma} X^\alpha X^\beta X^\gamma$.
They are contained in the following fuzzy analogues of the quadratic $CP^2$ constraints (\ref{cntuumvls}), 
 \beqa  X^\alpha X^\alpha |_{{\cal  H}_n} &=& \frac 43 + \frac 4n\label{fzyqdcsmr}\\ && \cr  d^{\alpha\beta\gamma} X^\alpha X^\beta |_{{\cal  H}_n}& =&\Bigl(\frac 13+\frac 1{2n}\Bigr) \; X^\gamma|_{{\cal  H}_n} \;, \eeqa 
 in addition to  \beqa   f^{\alpha\beta\gamma} X^\alpha X^\beta |_{{\cal  H}_n}& =&\frac{6i}n \; X^\gamma|_{{\cal  H}_n} \qquad\qquad\label{fXX}\eeqa 
  The quadratic constraints (\ref{fzyqdcsmr}-\ref{fXX}) tend towards the  commutative constraints  (\ref{cntuumvls})  in the large $n$ (or equivalently, large $N$) limit.
(\ref{fzyqdcsmr}) assigns a value to the quadratic Casimir, while for the cubic Casimir we then get
\be  d^{\alpha\beta\gamma} X^\alpha X^\beta X^\gamma |_{{\cal  H}_n} =\frac 49+\frac 2n +\frac 2{n^2}   \ee
The $CP^2_F$ algebra is the algebra of  $N\times N$ matrices which are polynomial functions of $X^\alpha$, satisfying the constraints  (\ref{fzyqdcsmr}-\ref{fXX}).  The standard choice for the Laplace operator on $CP^2_F$ is $\Delta_E=[X^\alpha,[X^\alpha,...]]$.

 Star products for  $CP^2_F$ are known.\cite{Balachandran:2001dd},\cite{Schneiderbauer:2016wub}  Using a star product, denoted by $\star$,  the  $CP^2_F$ algebra is mapped to a noncommutative algebra of  functions on  $CP^2$. So for example, from (\ref{criaib}), the images (or `symbols') ${\cal X}_\alpha$ of the operators  ${ X}_\alpha$ under the map  satisfy the star commutator:
\be
{\cal  X}^\alpha \star {\cal  X}^\beta - {\cal  X}^\beta \star{\cal  X}^\alpha  =\frac {2i}nf^{\alpha\beta\gamma}{\cal X}^{\gamma}\label{spcriaib}\ee
In the commutative limit $n\rightarrow \infty$, the star product of functions is required to reduce to the point-wise product (at zeroth order in $1/n$), and the star commutator of functions reduces to $i$ times the Poisson bracket of functions (at first order in $1/n$).  So for example, the left hand side of (\ref{spcriaib}) goes to $\frac in\{{\cal  X}^\alpha , {\cal  X}^\beta \}$ as $n\rightarrow \infty$, and in that limit, ${\cal  X}^\alpha $ satisfy the same Poisson bracket relations as $x^\alpha$ in (\ref{pbsfrmbdcrds}).  Therefore $\chi^\alpha$ can be identified with the $CP^2 $ embedding coordinates in the large $n$ limit.

\section{ $CP^2_F$ solutions to matrix models}
\setcounter{equation}{0}
\subsection{Euclidean background}

$CP^2_F$  is easily seen to be a solution of a Yang-Mills matrix model with a Euclidean background metric.  For this we introduce $M\times M$ matrices $Y^\alpha$, $\alpha=1,...,8$, whose dynamics is governed  by the action \cite{Azuma:2004qe}
\be  S_E(Y ) =\frac 1 {g^2}{\rm Tr}\Bigl(-\frac 14 [Y^\alpha,Y^\beta]^2 +\frac 23 i\tilde\alpha f^{\alpha\beta\gamma} Y^\alpha Y^\beta Y^\gamma\Bigr) \label{ukldnmact} \;,\ee where $\tilde\alpha$ is a real coefficient.  The first term in the trace defines the  Yang-Mills matrix action (which can be trivially extended to ten dimensions) appears in the IKKT matrix model.\cite{Ishibashi:1996xs}   It is invariant under rotations in the eight-dimensional Euclidean space.  This $SO(8)$ symmetry is broken by the second term, which instead  is invariant under the adjoint action of $SU(3)$, with infinitesimal variations $\delta Y^\alpha= 2i  f^{\alpha\beta\gamma} Y^\beta \epsilon^\gamma $, for infinitesimal parameters $\epsilon^\alpha$.  Both terms are invariant under the common subgroup of rotations in the $\alpha=1,2,3$ directions, as well as translations in the eight-dimensional Euclidean space. 

 The action (\ref{ukldnmact})  has extrema at
\be [[Y^\alpha, Y^\beta], Y^\beta] + i\tilde\alpha f^{\alpha\beta\gamma}[Y^\beta, Y^\gamma ]=0 \label{ucldneom}\ee
 $CP^2_F$ is a solution to  (\ref{ucldneom}).  That means we identify $ Y^\alpha$ with $N\times N$ matrix representations of the $ X^\alpha  $, defined in the previous section.
 For this we also need to make the identification $\tilde\alpha =2/n$, $n$ being an integer such that   $N=\frac{ (n+2)!}{
2 n!}\le M\;$.

\subsection{Lorentzian  background}

The matrix action (\ref{ukldnmact}) was written in an
eight-dimensional Euclidean ambient space.
Here we change the ambient space to eight-dimensional Minkowski space, with metric tensor $\eta={\rm diag}(1,1,1,1,1,1,1,-1)$.   In order to find nontrivial solutions we also add a quadratic term to the action, which now reads
\be  S_M(Y ) =\frac 1 {g^2}{\rm Tr}\Bigl(-\frac 14 [Y^\alpha,Y^\beta] [Y_\alpha,Y_\beta] +\frac 23 i\tilde\alpha f^{\alpha\beta\gamma} Y_\alpha Y_\beta Y_\gamma+\frac\beta 2Y^\alpha Y_\alpha\Bigr) \label{mnkact} \;,\ee 
where $\beta$ is real and indices raised and lowered using $\eta$.  The action is an extremum when
 \be [[Y^\alpha, Y^\beta], Y_\beta] + i\tilde\alpha f^{\alpha\beta\gamma}[Y_\beta, Y_\gamma ]+\beta Y^\alpha =0\label{four.4}\ee

A simple solution $Y^\alpha =\bar Y^\alpha$ to (\ref{four.4})  is   $CP^2_F$, now written  in a  Lorentzian background:
\be \bar Y^\alpha=n\tilde\alpha\, X^\alpha\;\label{fzeelcp2}\ee
Here $\tilde\alpha$ and $\beta$  are constrained by
\be \beta = - 6 \tilde\alpha^2\label{alfbta} \ee
 For any fixed $n$, which defines a matrix representation, this solution is expressed in terms of only one free parameter, which sets an overall scale.  This $CP^2$ solution is not invariant under all of  $SU(3)$, since general transformations do not preserve the time-like direction of the background metric.  On the other hand, the time-like direction is preserved under the adjoint action of the $SU(2)\times U(1)$ subgroup. In order for the Laplace operator associated with this solution to be consistent with the  eight-dimensional Minkowski metric tensor $\eta$, we should take it  to be $\Delta_M=[Y^\alpha,[Y_\alpha,...]]$, rather than the standard Laplace operator on $CP_F^2$.

A more general solution to (\ref{four.4}) which is also invariant under $SU(2)\times U(1)$  is
\beqa \bar Y^i &=& \;\; \frac {n\rho}2 \,  X^i \;,\qquad  i=1,2,3\cr &&\cr
\bar Y^a &=&v\, \frac {n\rho}2 \,  X^a \;,\qquad  a=4,5,6,7\cr &&\cr
\bar Y^8 &=&w\,\frac {n\rho}2  \,  X^8\;\,\quad \label{drmdlfcp2}\eeqa 
where the parameters $v,w,\rho,\tilde\alpha$ and $\beta$ are constrained by
\beqa  v&=&\frac{1}{2} \sqrt{\frac{\gamma +5 + w- w^2-w^3}{1+w}}\cr&&\cr
\frac{\tilde\alpha}\rho &=&\frac{5 + w+7  w^2-w^3-\gamma }{4 \left(1+4  w-w^2\right)} \cr&&\cr 
\frac\beta{\rho^2} &=&-\frac{3 \left(1+15  w -8 w^3 
   -w^4+w^5+\left(1+ 2w - w^2\right)\gamma \right)}{4 (1+w) \left(1+4  w-w^2\right)}
  \label{2prmtrsln}
   \eeqa and 
   \be\gamma =\sqrt{25 -6  w+7  w^2+4  w^3-17  w^4+2 
   w^5+w^6}\label{gamma}\ee For any fixed $n$, this solution is determined by two parameters $\rho$ and $w$, the former of which sets the overall scale.  Again, here we assume the Laplace operator to be $\Delta_M=[Y^\alpha,[Y_\alpha,...]]$. The solution is a one-parameter deformation of the previous $CP^2_F$ solution, given by (\ref{fzeelcp2}) and (\ref{alfbta}), and we can regard $w$ as the deformation parameter.  The previous solution is recovered for $w=1$, since then (\ref{2prmtrsln}) gives $v=1$,  $ \tilde\alpha=\frac \rho 2$ and $\beta=-\frac 32 \rho^2$. $v$ is real and finite for the domain   $-1< w\,
{}_\sim^<\, 1.32247$. $v$ tends towards the lower bound  $\approx.493295$ as $w$ goes to the upper limit $\approx 1.32247$, while $v$  is singular in the limit  $w\rightarrow -1$.  $v$ versus $w$ is plotted for this domain in figure 1.    
\begin{figure}[placement h]
\centering
  \includegraphics[height=2.in,width=2.in,angle=0]{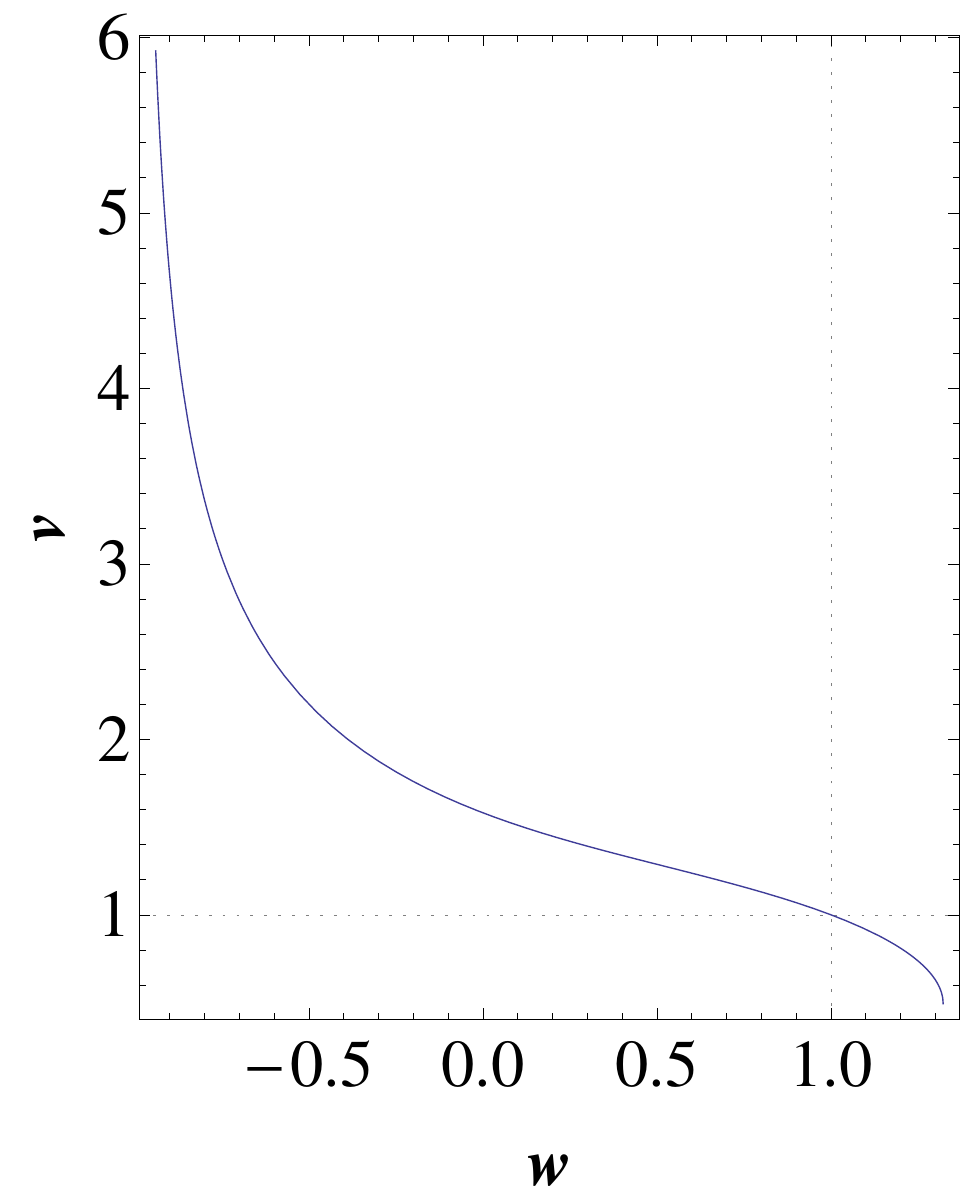}
\caption {$v$  versus $w$ is plotted for the one-parameter family of deformed $CP^2_F$ solutions given in (\ref{drmdlfcp2}) and (\ref{2prmtrsln}).}  
\label{fig:test}
\end{figure}

\section{Commutative limit}
\setcounter{equation}{0}

The discussions in section  four assume $N\times N$ matrix representations for  the  $CP^2_F$ solution  (\ref{fzeelcp2}), (\ref{alfbta}), and  the deformed    $CP^2_F$ solution (\ref{drmdlfcp2})-(\ref{gamma}).  Here we take the   $N\rightarrow \infty$ limit of these solutions  to reveal different space-time manifolds.  We begin with the undeformed $CP^2_F$ solution  (\ref{fzeelcp2}), (\ref{alfbta}).  

\subsection{ $CP^2$   in a  Lorentzian background}
For convenience we first fix the scale of the solution   (\ref{fzeelcp2}), (\ref{alfbta}) by setting $\tilde \alpha =1/n$ and $\beta=-1/n^2$.  Thus $\bar Y^\alpha=X^\alpha$, for any  $n$.  For some  star product we can introduce their corresponding symbols $ \bar {\cal Y}^\alpha$, and  so   $ \bar {\cal Y}^\alpha=  {\cal X}^\alpha$, where $ {\cal X}^\alpha$ satisfies the star commutator (\ref{spcriaib}).  Then in the  $n\rightarrow \infty$ limit,  $ \bar {\cal Y}^\alpha$ obey the Poisson brackets (\ref{pbsfrmbdcrds}),  and constraints of the form (\ref{cntuumvls}).  In the limit, $ \bar {\cal Y}^\alpha$ can  be expressed in terms of complex coordinates $z$ as in 
 (\ref{cmsu3gen}), which once again span a four-dimensional manifold.  However, now the manifold, strictly speaking, is  not $CP^2$.  While we recover the $CP^2$ constraints  (\ref{cntuumvls}) and (\ref{pbsfrmbdcrds}) in the commutative  limit, the induced metric on the manifold cannot be the Fubini-Study metric (\ref{FSm}).  The latter followed from the Euclidean background metric tensor on ${\mathbb{R}^8}$, given in (\ref{atedemtrc}).  Now the embedding matrices $ \bar { Y}^\alpha$, and their symbols $ \bar {\cal Y}^\alpha$, span eight-dimensional  Minkowski space.  Moreover,   since the Laplace operator for the matrix solution (\ref{fzeelcp2}) is constructed using the eight-dimensional Minkowski metric tensor $\eta$, the  induced metric tensor on the surface that is recovered in  the $n\rightarrow \infty$ limit of the solution should also  be constructed using $\eta$.  The induced metric tensor on the surface is thus computed from the invariant  length for  the eight-dimensional Minkowski  space,
\be ds_M^2= d \bar {\cal Y}^\alpha d \bar {\cal Y}_\alpha= ds_E^2- 2( d  {\cal X}^8)^2  \;,\ee
where we assume  $ \bar {\cal Y}^\alpha=  {\cal X}^\alpha$.
Then by writing $ {\cal X}^\alpha=\frac {\bar z \lambda^\alpha z} {|z|^2}$, one gets corrections to the Fubini-Study metric
\be ds_M^2 = ds_E^2\;-\ \frac{2\,\Bigl(d(\bar z \lambda^8 z)\Bigr)^2}{{|z|^4}}\;  -\; \frac{2\,(\bar z \lambda^8 z)^2\,\Bigl(d|z|^2\Bigr)^2-d|z|^4\,d(\bar z \lambda^8 z)^2}{{|z|^8}}    \ee
In terms of the   coordinates
 $\zeta_a= z_a/z_3$, $a=1,2$,  which are invariant under $z\rightarrow \gamma z$, we  get
\beqa ds_M^2&= &ds_E^2\;-\ \frac{24 |\zeta|^2}{(|\zeta|^2+1)^4}\,(d|\zeta|)^2\cr &&\cr &=& 4\,\frac{ (|\zeta|^2-1)^2-2|\zeta|^2}{(|\zeta|^2+1)^4}\,(d|\zeta|)^2   +\frac{4 |\zeta|^2}{(|\zeta|^2+1)}\,(\omega_1^2+\omega_2^2)+\frac{4 |\zeta|^2}{(|\zeta|^2+1)^2}\,\omega_3^2\;,\label{three2one}\eeqa
where the left-invariant one forms $\omega_i$ were defined previously in section two.

 The  metric tensor obtained here differs from that on $CP^2$, and furthermore  is  not  K\"ahler.  On the other hand, the symplectic two-form remains unchanged, i.e. it is  (\ref{smplct2frm3}).   $SU(3)/Z_3$ is no longer an isometry.  Instead,  the metric tensor (\ref{three2one}) and   symplectic two-form  are invariant under $SU(2)\times U(1)/Z_2$, generated by the Killing vectors (\ref{klngvctrs}).   A novel feature is that the metric tensor has variable signature.  It has Euclidean signature for $0<|\zeta|^2<2-\sqrt{3}$ and $|\zeta|^2>2+\sqrt{3}$, and  Lorentzian signature for $2-\sqrt{3}<|\zeta|^2<2+\sqrt{3}$. 
The metric tensor, along with the Ricci scalar, is singular at the boundaries $|\zeta|^2=2\pm\sqrt{3}$ between the regions, and so the boundaries define physical singularities.  [There are also coordinate singularities  located at
$|\zeta|=0$ and $|\zeta|\rightarrow\infty$, just as is the case with the $CP^2$ metric tensor given by (\ref{cp2mtrcomga}).] 
Away from the singularities, the manifold is spatially homogeneous and axially symmetric at each point, and the invariant length (\ref{three2one}) has a form which is similar to that of a Taub-NUT space (more specifically, the Taub region of Taub-NUT space since the coefficient of $\omega_3^2$ is positive).

We now   restrict to the Lorentzian region $2-\sqrt{3}<|\zeta|^2<2+\sqrt{3}$.  $|\zeta|$ is  a time parameter in this region, and one has the following properties:

\noindent
a) There are time-like geodesics which originate at the initial singularity, which we choose to be at $|\zeta|=\sqrt{2-\sqrt{3}}$, and terminate at the final singularity at $|\zeta|=\sqrt{2+\sqrt{3}}$.  The elapsed  proper time along  a geodesic with  $\omega_1=\omega_2=\omega_3=0$ can be written as a function of $|\zeta|$
\be \tau(|\zeta|)= 2\int_{\sqrt{2-\sqrt{3}}}^{|\zeta|}\;\frac{\sqrt{-r^4+4r^2-1}}{(r^2+1)^2}\; dr \label{prprtmongeo}\ee
The the total proper time from the initial singularity to the final singularity is $\tau\Bigl(\sqrt{2+\sqrt{3}}\Bigr)\approx .672\,. $

\noindent 
b) From the volume of any time-slice, which can be constructed from  the determinant of the metric, ${}^3g|_{|\zeta |}$,   on a time-slice, one can assign a spatial distance scale $a$ as a function of $|\zeta|$,
\be a(|\zeta|)^3= \int\sqrt{{}^3g|_{|\zeta |}}\;d\theta d\phi d\psi= \frac{8\pi^2 |\zeta|^3}{(|\zeta|^2+1)^2}\;,\label{tmslcvlm}\ee
where  the integration is done on the  time-slice,  which can be parametrized by the Euler angles in (\ref{Oilerngls}).  A novel feature of this space-time is that the distance scale is  nonvanishing at the time of the initial and final singularities, corresponding to $|\zeta|=\sqrt{2-\sqrt{3}}$ and  $|\zeta|=\sqrt{2+\sqrt{3}}$, respectively, 
$$  a(\sqrt{2-\sqrt{3}})\approx 1.896 \qquad\quad a(\sqrt{2+\sqrt{3}})\approx 2.940 $$
A plot  of   the normalized scale
$a/a|_{\tau=0}$ as a function of the time $\tau$ from  $\tau=0$ (the time of the initial singularity) to the time of the final singularity appears in  in figure 3 (solid curve). It  is  seen to grow and   de-accelerate.

\subsection{ Deformed $CP^2$   in a  Lorentzian background} 

We can obtain a one-parameter family of space-time manifolds, including the one obtained in the above subsection, by taking the commutative limit of  the   deformed    $CP^2_F$ solution (\ref{drmdlfcp2})-(\ref{gamma}).
Here it is convenient to  set $\rho=2/n$.   Then the  symbols $\bar {\cal Y}^\alpha$ of the matrices $\bar Y^\alpha$ for the solution in (\ref{drmdlfcp2})  satisfy
\beqa \bar {\cal Y}^i &=& \;\; \, {\cal X}^i \;,\qquad  i=1,2,3\cr &&\cr
\bar {\cal Y}^a &=&v\,{\cal  X}^a \;,\qquad  a=4,5,6,7\cr &&\cr
\bar {\cal Y}^8 &=&w\,{\cal X}^8\;\,\quad \label{smbldrmdlfcp2}\;,\eeqa 
where ${\cal X}^\alpha$ again denote the symbols of the  $CP^2_F$ matrices. Recall  $v$ is real and finite for the domain   $-1< w\,
{}_\sim^<\, 1.32247$, while $w$ is given in (\ref{2prmtrsln}) and plotted in figure 1.  In the $n\rightarrow \infty$ limit, we shall keep $v$ and $w$ fixed, which implies as before that $\tilde\alpha$ and $\beta$ vanish in the limit, $\tilde\alpha\sim 1/n$ and $\beta\sim 1/n^2$. 
 The invariant  length in the eight-dimensional Minkowski space now reads
\be ds_M^2=d \bar {\cal Y}^\alpha d \bar {\cal Y}_\alpha= v^2 ds_E^2 +(1-v^2)( d  {\cal X}^i)^2 - (w^2+v^2)( d  {\cal X}^8)^2\;,\ee  where we substituted the commutative solution (\ref{smbldrmdlfcp2}).
Using the identities 
\be ( d  {\cal X}^i)^2 =\frac{4\,|\zeta|^4}{(1+|\zeta|^2)^2}\,(\omega^2_1+\omega^2_2)+   \frac {4\,|\zeta|^2}{(1+|\zeta|^2)^4}(d|\zeta|)^2\qquad\qquad  ( d  {\cal X}^8)^2 =  \frac {12\,|\zeta|^2}{(1+|\zeta|^2)^4}(d|\zeta|)^2\;,\ee 
which follows from  $ {\cal X}^\alpha=\frac {\bar z \lambda^\alpha z} {|z|^2}$ and the previous definition of the left-invariant one forms $\omega_i$, we now get
\be ds_M^2= 4\biggl(\frac {v^2(|\zeta|^2-1)^2+(1-3w^2)|\zeta|^2}{(1+|\zeta|^2)^4}\biggr)(d|\zeta|)^2  +\frac{4|\zeta|^2 (v^2+|\zeta|^2)}{(1+|\zeta|^2)^2}\,(\omega^2_1+\omega^2_2)+\frac{4\, v^2|\zeta|^2}{(|\zeta|^2+1)^2}\,\omega_3^2\;\label{gnrlmt}\ee 
This expression reduces to (\ref{three2one}) when $w=v=1$.
The symplectic two-form is again given by  (\ref{smplct2frm3}).

As in the previous case,
 the   metric tensor and   symplectic two-form  are invariant under $SU(2)\times U(1)/Z_2$, generated by the Killing vectors (\ref{klngvctrs}). 
The induced metric tensor now has physical singularities at $|\zeta|=|\zeta_\pm|$, where 
\be |\zeta_\pm|^2=\frac{2v^2+3w^2-1\pm\sqrt{(3w^2-1)(4v^2+3w^2-1)}}{2v^2} \;,\label{sngnzta}\ee  which using (\ref{2prmtrsln}) are functions of only $w$.  The singularities are plotted as a function of $w$ in figure 2.   There are two singularities for  the domains   $-1 >w >-  \frac 1{\sqrt{3}}$ and  $ \frac 1{\sqrt{3}}< w\,
{}_\sim^<\, 1.32247$,  one singularity (at $|\zeta|=1$)  for $w=\pm\frac 1{\sqrt{3}}$, and none for $-  \frac 1{\sqrt{3}}< w< \frac 1{\sqrt{3}}$. As before, they define the boundaries between regions of Euclidean signature and Lorentzian signature. (The  regions of  Lorentzian signature are  shaded  in the figure.) For the domain  $-  \frac 1{\sqrt{3}}< w< \frac 1{\sqrt{3}}$,   the metric tensor in (\ref{gnrlmt}) has a  Euclidean signature for all $|\zeta|^2$.
\begin{figure}[placement h]
\centering
  \includegraphics[height=2.25in,width=2.in,angle=0]{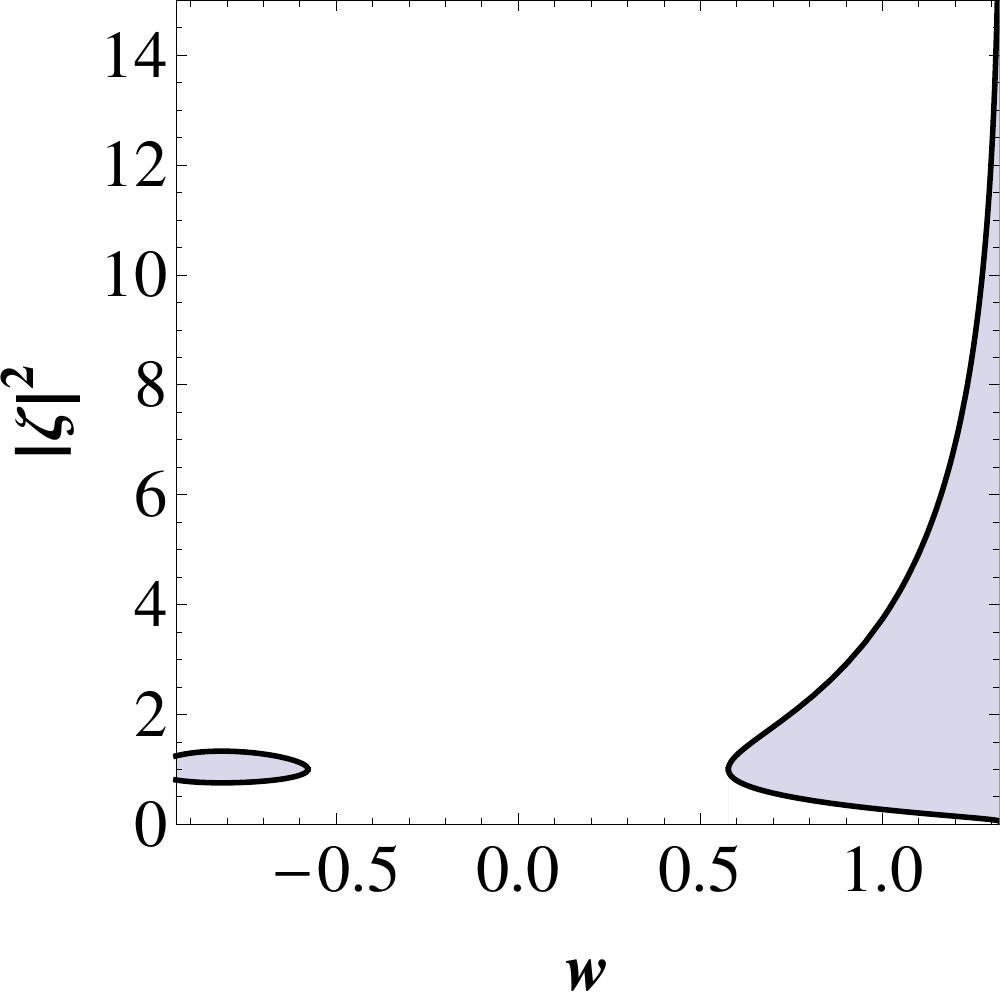}
\caption {Singularities at  $|\zeta|^2=|\zeta_\pm|^2$ are plotted as a function of $w$ using (\ref{sngnzta}).  They define boundarises between regions of Lorentzian signature (shaded) and Euclidean signature (unshaded).}  
\label{fig:test}
\end{figure}

 Once again there  are time-like geodesics which originate at the initial singularity, which we choose to be at $|\zeta|=|\zeta_-|$, and terminate at the final singularity at  $|\zeta|=|\zeta_+|$. The generalization of the expression (\ref{prprtmongeo}) for the elapsed  proper time along  a geodesic with  $\omega_1=\omega_2=\omega_3=0$ can be written as
\be \tau(|\zeta|)= 2\int_{|\zeta_-|}^{|\zeta|}\;\frac{\sqrt{(3w^2-1)r^2-v^2(r^2-1)^2}}{(r^2+1)^2}\; dr \label{prprtmongeognrl}\ee
 The generalization of the expression   (\ref{tmslcvlm}) for the volume of a $|\zeta|$-slice, which we again denote by  $a(|\zeta|)^3$, is 
\be a(|\zeta|)^3= \int\sqrt{{}^3g|_{|\zeta |}}\;d\theta d\phi d\psi= \frac{8\pi^2 v |\zeta|^3(|\zeta|^2+v^2)}{(|\zeta|^2+1)^3}\;,\label{tmslcvlmgnl}\ee

 We restrict to the  region of Lorentzian signature for four different choices for $w$ (and hence $v$), including the case $w=v=1$ of the previous subsection, in figure 3. There we plot the normalized scale $a/a|_{\tau=0}$  as a function of the time $\tau$, starting from  $\tau=0$ (the time of the initial singularity) to the time of the final singularity. In all cases the distance scale $a$ is  nonvanishing at the time of the initial and final singularities, and the scale grows and   de-accelerates.
 The largest and longest  expansion occurs when $w$ takes its maximum value of $\sim 1.3225$, while the space-time only exists for an instant for  $w=\pm\frac 1{\sqrt{3}}$.
\begin{figure}[placement h]
\centering
  \includegraphics[height=2.75in,width=2.5in,angle=0]{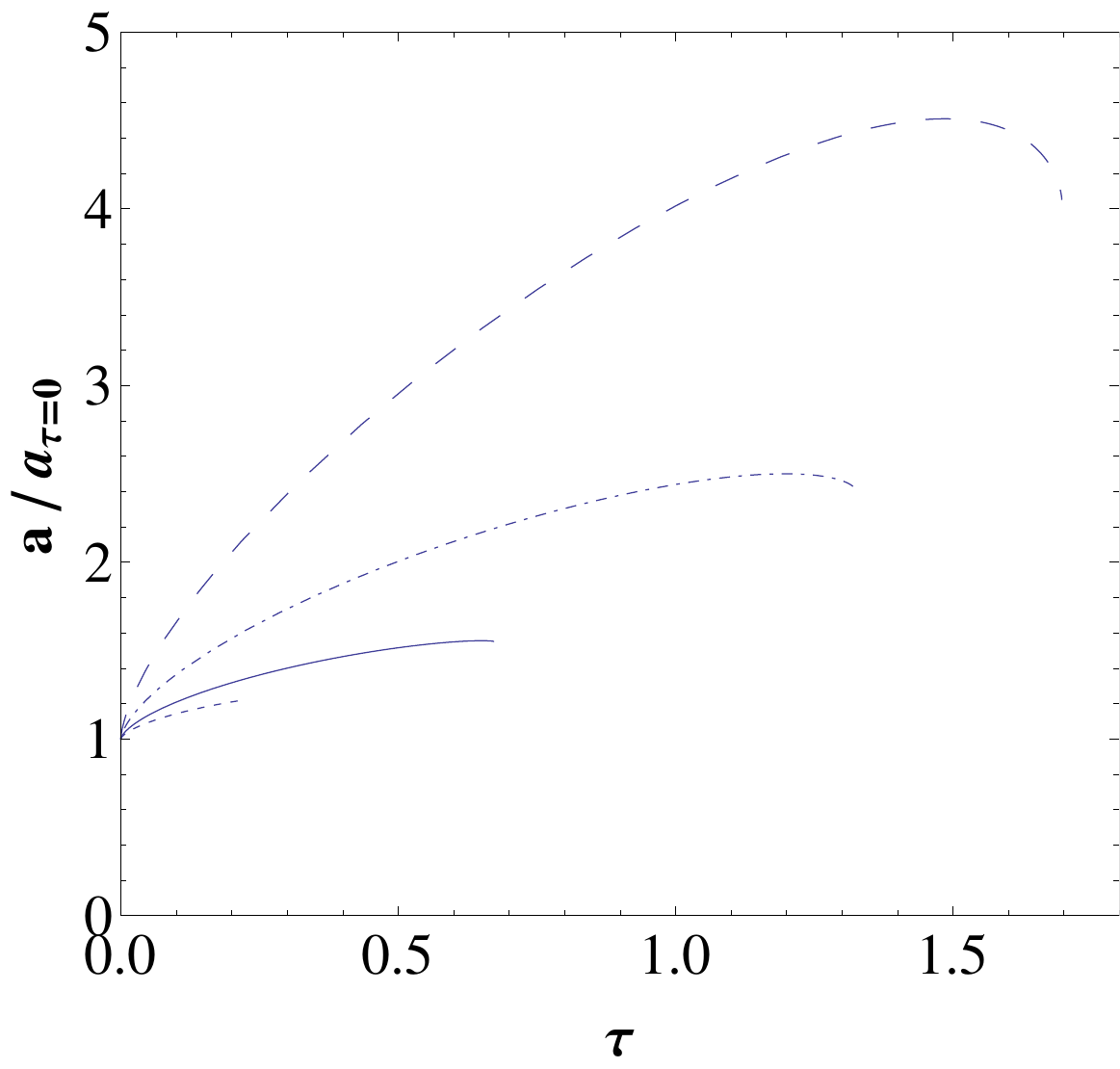}
\caption {$a/a|_{\tau=0}$ as a function of the time $\tau$ from  $\tau=0$ (the time of the initial singularity) to the time of the final singularity for four different choices for $w$ (and hence $v$): $w\approx 1.3225$ (large dashed curve), $w=1.25$ (dot-dashed curve), $w=1$ (solid curve) and $w=.75$ (small dashed curve). }  
\label{fig:test}
\end{figure}

\section{Concluding remarks}
\setcounter{equation}{0}

We have constructed four-dimensional manifolds by taking the $N\rightarrow \infty$ of solutions to Lorentzian matrix equations   (\ref{mnkact}). The metric tensor and  symplectic two-form on the manifold are invariant under $SU(2)\times U(1)/Z_2$. The manifolds, in general, have  changing signature.  We get toy  models of space-time after restricting to regions with Lorentzian signature, complete with initial and final cosmological singularities. The metric tensor resembles that of  the Taub region of Taub-NUT space.  In all cases,  the distance scale scale grows and   de-accelerates as shown in  figure 3, which clearly does not give a realistic picture of our universe.

Many other solutions of the Lorentzian matrix equations (\ref{mnkact}) are  possible. On the other hand, not all solutions may  have a well defined commutative  (or large $N$) limit. One such example is
\beqa \bar Y^i &=&  -n\,(2+\sqrt{5})\,  X^i \;,\qquad  i=1,2,3\cr &&\cr 
\bar Y^a &=&  -n\,\sqrt{ 29+13\sqrt{5}\, }\;  X^a \;,\qquad  a=4,5,6,7\cr &&\cr
\bar Y^8 &=& n\,  X^8\;\,\quad \eeqa where again  $ X^\alpha$ is defined in (\ref{dffzeecp2}). 
In this case both $\tilde\alpha$ and $\beta$ are fixed,  $
 \tilde\alpha=\frac 2{19}(10-\sqrt{5}) $, $  \beta=-\frac{24}{19}(163+73\sqrt{5})$.
Again, the dimension of the representation is  $N=\frac{ (n+2)!}{
2 n!}\;.$ Now the only free parameter is $n$, and the solution is ill-defined when $n\rightarrow \infty$ and so there is no commutative limit.  Upon modifying the matrix action  (\ref{mnkact}), in particular the cubic term, it should be possible to find  solutions associated with other noncommutative geometries, which may or may not have a commutative limit.  One possibility is the fuzzy four-sphere embedded in a Lorenzian background.

Many other issues can  be explored.   Among them are: the question of  stability for the various classical matrix solutions, the role played by the inclusion of fermionic degrees of freedom in the matrix model, and the computation of  quantum effects.  With regard to fermions, we note that supersymmetry, in addition to translation symmetry, is explicitly broken by the presence of the quadratic term in the action (\ref{mnkact}).
Of course, it is also of interest to investigate whether a more physical cosmology can be found amongst  the solutions of this, or related, matrix models.  Since a compact coset space necessarily implies a closed  space-time cosmology,  to get an open universe one proposal is to start with a noncompact noncommutative coset space.  One expects matrix representations then to be infinite-dimensional, and although one cannot then take $N\rightarrow \infty$, it should be possible to define an alternative commutative limit  in this case.  
A striking feature of the space-times recovered in section five is that initial singularity occurs when the universe has a nonzero distance scale $a(|\zeta_-|)$. This distance scale should be greater than the Planck length since Planck scale effects are washed out in the continuum limit.  Time cannot be defined for distance scales smaller than $a(|\zeta_-|)$.  If this feature, i.e., that the universe begins with a non zero spatial size, can be implemented in a realistic cosmology, then  it may not be necessary to consider the very early universe, and perhaps,  one can even avoid having an inflationary era.

\bigskip
{\Large {\bf Acknowledgments} }

\noindent
We are very grateful to A.  Pinzul for valuable discussions.


\bigskip

\begin{thebibliography}{99}

\bibitem{Madore:1991bw} 
  J.~Madore,
  ``The fuzzy sphere,''
  Class.\ Quant.\ Grav.\  {\bf 9}, 69 (1992).

\bibitem{Grosse:1994ed} 
  H.~Grosse and P.~Presnajder,
  ``The Dirac operator on the fuzzy sphere,''
  Lett.\ Math.\ Phys.\  {\bf 33}, 171 (1995).

\bibitem{CarowWatamura:1998jn} 
  U.~Carow-Watamura and S.~Watamura,
  ``Noncommutative geometry and gauge theory on fuzzy sphere,''
  Commun.\ Math.\ Phys.\  {\bf 212}, 395 (2000).

\bibitem{Alexanian:2000uz} 
  G.~Alexanian, A.~Pinzul and A.~Stern,
  ``Generalized coherent state approach to star products and applications to the fuzzy sphere,''
  Nucl.\ Phys.\ B {\bf 600}, 531 (2001).

\bibitem{Dolan:2001gn}  
  B.~P.~Dolan, D.~O'Connor and P.~Presnajder,
  ``Matrix $\phi^4$ models on the fuzzy sphere and their continuum limits,''
  JHEP {\bf 0203}, 013 (2002).

\bibitem{Balachandran:2002jf} 
  A.~P.~Balachandran, S.~Kurkcuoglu and E.~Rojas,
  ``The star product on the fuzzy supersphere,''
  JHEP {\bf 0207}, 056 (2002)


\bibitem{Balachandran:2005ew} 
  A.~P.~Balachandran, S.~Kurkcuoglu and S.~Vaidya,
  ``Lectures on fuzzy and fuzzy SUSY physics,''
  Singapore, Singapore: World Scientific (2007) 191 p. [hep-th/0511114]. 


\bibitem{Iso:2001mg} 
  S.~Iso, Y.~Kimura, K.~Tanaka and K.~Wakatsuki,
  ``Noncommutative gauge theory on fuzzy sphere from matrix model,''
  Nucl.\ Phys.\ B {\bf 604}, 121 (2001).



\bibitem{Chaney:2015mfa} 
  A.~Chaney, L.~Lu and A.~Stern,
 ``Lorentzian Fuzzy Spheres,''
  Phys.\ Rev.\ D {\bf 92}, no. 6, 064021 (2015);
``Matrix Model Approach to Cosmology,''
  Phys.\ Rev.\ D {\bf 93}, no. 6, 064074 (2016). 

\bibitem{Alvarez:1997fy} 
  E.~Alvarez and P.~Meessen,
  ``Newtonian M(atrix) cosmology,''
  Phys.\ Lett.\ B {\bf 426}, 282 (1998).

\bibitem{Freedman:2004xg} 
  D.~Z.~Freedman, G.~W.~Gibbons and M.~Schnabl,
  ``Matrix cosmology,''
  AIP Conf.\ Proc.\  {\bf 743}, 286 (2005).

\bibitem{Craps:2005wd} 
  B.~Craps, S.~Sethi and E.~P.~Verlinde,
  ``A Matrix big bang,''
  JHEP {\bf 0510}, 005 (2005).


\bibitem{Erdmenger:2007xs} 
  J.~Erdmenger, R.~Meyer and J.~H.~Park,
  ``Spacetime Emergence in the Robertson-Walker Universe from a Matrix model,''
  Phys.\ Rev.\ Lett.\  {\bf 98}, 261301 (2007).

\bibitem{Klammer:2009ku}
  D.~Klammer and H.~Steinacker,
  ``Cosmological solutions of emergent noncommutative gravity,''
  Phys.\ Rev.\ Lett.\  {\bf 102} (2009) 221301.

\bibitem{Kim:2011ts} 
 S.~W.~Kim, J.~Nishimura and A.~Tsuchiya,
 ``Expanding (3+1)-dimensional universe from a Lorentzian matrix model for superstring theory in (9+1)-dimensions,''
  Phys.\ Rev.\ Lett.\  {\bf 108}, 011601 (2012); S.~W.~Kim, J.~Nishimura and A.~Tsuchiya, ``Expanding universe as a classical solution in the Lorentzian matrix model for nonperturbative superstring theory,''
  Phys.\ Rev.\ D {\bf 86}, 027901 (2012); 
``Late time behaviors of the expanding universe in the IIB matrix model,''
  JHEP {\bf 1210}, 147 (2012);
  Y.~Ito, S.~W.~Kim, J.~Nishimura and A.~Tsuchiya,
  ``Monte Carlo studies on the expanding behavior of the early universe in the Lorentzian type IIB matrix model,''
  PoS LATTICE {\bf 2013}, 341 (2014);  Y.~Ito, S.~W.~Kim, Y.~Koizuka, J.~Nishimura and A.~Tsuchiya,
 ``A renormalization group method for studying the early universe in the Lorentzian IIB matrix model,''
  PTEP {\bf 2014}, no. 8, 083B01 (2014).






\bibitem{Arnlind:2012cx}J.~Arnlind and J.~Hoppe,
``The world as quantized minimal surfaces,''
Phys.\ Lett.\ B {\bf 723}, 397 (2013).

\bibitem{Jurman:2013ota}
D.~Jurman and H.~Steinacker,
``2D fuzzy Anti-de Sitter space from matrix models,'' JHEP {\bf 1401}, 100 (2014).





\bibitem{Stern:2014aqa} A.~Stern,
``Noncommutative Static Strings from Matrix Models,''
Phys.\ Rev.\ D {\bf 89}, no. 10, 104051 (2014); 
``Matrix Model Cosmology in Two Space-time Dimensions,''
Phys.\ Rev.\ D {\bf 90}, no. 12, 124056 (2014).



\bibitem{Sakharov:1984ir} 
  A.~D.~Sakharov,
  ``Cosmological Transitions With a Change in Metric Signature,''
  Sov.\ Phys.\ JETP {\bf 60}, 214 (1984)
  [Zh.\ Eksp.\ Teor.\ Fiz.\  {\bf 87}, 375 (1984)]
  [Sov.\ Phys.\ Usp.\  {\bf 34}, 409 (1991)].
  
 
\bibitem{Gibbons:1990ns} 
  G.~W.~Gibbons and J.~B.~Hartle,
  ``Real Tunneling Geometries and the Large Scale Topology of the Universe,''
  Phys.\ Rev.\ D {\bf 42}, 2458 (1990).
 
   
\bibitem{Dray:1991zz} 
  T.~Dray, C.~A.~Manogue and R.~W.~Tucker,
  ``Particle production from signature change,''
  Gen.\ Rel.\ Grav.\  {\bf 23}, 967 (1991).  
  
\bibitem{Ellis:1991st}
  G.~Ellis, A.~Sumeruk, D.~Coule and C.~Hellaby,
  ``Change of signature in classical relativity,''
  Class.\ Quant.\ Grav.\  {\bf 9} (1992) 1535.  

\bibitem{Hayward:1992zp} 
  S.~A.~Hayward,
  ``Signature change in general relativity,''
  Class.\ Quant.\ Grav.\  {\bf 9}, 1851 (1992). 
   

\bibitem{Mars:2000gu} 
  M.~Mars, J.~M.~M.~Senovilla and R.~Vera,
 ``Signature change on the brane,''
  Phys.\ Rev.\ Lett.\  {\bf 86}, 4219 (2001).
  
\bibitem{Mielczarek:2012pf} 
  J.~Mielczarek,
  ``Signature change in loop quantum cosmology,''
  Springer Proc.\ Phys.\  {\bf 157}, 555 (2014).

\bibitem{Barrau:2016sqp} 
  A.~Barrau and J.~Grain,
  ``Cosmology without time: What to do with a possible signature change from quantum gravitational origin?,''
  arXiv:1607.07589 [gr-qc].

\bibitem{Bojowald:2016vlj} 
  M.~Bojowald and S.~Brahma,
  ``Signature change in loop quantum gravity: General midisuperspace models and dilaton gravity,''
  arXiv:1610.08840 [gr-qc].

\bibitem{Ambjorn:2015qja} 
  J.~Ambjorn, D.~N.~Coumbe, J.~Gizbert-Studnicki and J.~Jurkiewicz,
 ``Signature Change of the Metric in CDT Quantum Gravity?,''
  JHEP {\bf 1508}, 033 (2015).
  
 \bibitem{Ishibashi:1996xs} 
  N.~Ishibashi, H.~Kawai, Y.~Kitazawa and A.~Tsuchiya,
  ``A Large N reduced model as superstring,''
  Nucl.\ Phys.\ B {\bf 498}, 467 (1997). 

\bibitem{Alexanian:2001qj}
  G.~Alexanian, A.~P.~Balachandran, G.~Immirzi and B.~Ydri,
  ``Fuzzy CP**2,''
  J.\ Geom.\ Phys.\  {\bf 42} (2002) 28.
  

\bibitem{Azuma:2004qe} 
  T.~Azuma, S.~Bal, K.~Nagao and J.~Nishimura,
  ``Dynamical aspects of the fuzzy CP**2 in the large N reduced model with a cubic term,''
  JHEP {\bf 0605}, 061 (2006).  
\bibitem{Grosse:2004wm} 
  H.~Grosse and H.~Steinacker,
  ``Finite gauge theory on fuzzy CP**2,''
  Nucl.\ Phys.\ B {\bf 707}, 145 (2005).  
   
\bibitem{Janssen:2004cd} 
  B.~Janssen, Y.~Lozano and D.~Rodriguez-Gomez,
  ``Giant gravitons and fuzzy CP**2,''
  Nucl.\ Phys.\ B {\bf 712}, 371 (2005). 
  
 

  \bibitem{Dou:2007in} 
  D.~Dou and B.~Ydri,
  ``Topology change from quantum instability of gauge theory on fuzzy CP**2,''
  Nucl.\ Phys.\ B {\bf 771}, 167 (2007).
  

  \bibitem{Ydri:2016dmy} 
  B.~Ydri,
  ``Lectures on Matrix Field Theory I,''
  Lect.\ Notes Phys.\  {\bf 929}, pp. (2017).

  

  
\bibitem{Balachandran:2001dd} 
  A.~P.~Balachandran, B.~P.~Dolan, J.~H.~Lee, X.~Martin and D.~O'Connor,
  ``Fuzzy complex projective spaces and their star products,''
  J.\ Geom.\ Phys.\  {\bf 43}, 184 (2002).

 \bibitem{Schneiderbauer:2016wub} 
  L.~Schneiderbauer and H.~C.~Steinacker,
  ``Measuring finite Quantum Geometries via Quasi-Coherent States,''
  J.\ Phys.\ A {\bf 49}, no. 28, 285301 (2016). 

  
\end{thebibliography}
\end{document}